\documentclass[twocolumn,showpacs,preprintnumbers,amsmath,amssymb,prb]{revtex4}

\usepackage{graphicx}
\usepackage{dcolumn}
\usepackage{bm}
\usepackage{amsmath}

\begin{document}

\title{$^{55}$Mn NMR and magnetization studies of La$_{0.67}$Sr$_{0.33}$MnO$_3$ thin films}

\author{A. A. Sidorenko}
\email{sidorenko@fis.unipr.it}
\author{G. Allodi}
\author{R. De Renzi}
\affiliation{Dipartimento di Fisica e Unit\`a CNISM di Parma, Universit\`a degli Studi di Parma, Viale delle Scienze, 7A, 43100 Parma, Italy}
\author{G. Balestrino}
\author{M. Angeloni}
\affiliation{INFM-COHERENTIA, Dipartimento di Ingegneria Meccanica, Universit\`a di Roma Tor Vergata, Via del Politecnico, 1, 00133 Roma, Italy}
\date{\today}

\begin{abstract}
$^{55}$Mn nuclear magnetic resonance and magnetization studies of the series of  La$_{0.67}$Sr$_{0.33}$MnO$_3$ thin films have been performed at low temperature. Two distinct lines  were observed, at 322 MHz and 380 MHz, corresponding to two different phases, the former located at the interface, with localized charges, and the latter corresponding to the film bulk, with itinerant carriers (as it was also found in Ca manganite films). 
The spin-echo amplitude was measured as a function of a dc magnetic field applied either in the film plane or perpendicular to it. The field dependence of both the main NMR signal intensity and frequency shift is quite consistent with that calculated in a simple single domain model. The best fit to the model shows that magnetization rotation processes play a dominant role when the applied field exceeds the effective anisotropy field. Distinctly different magnetic anisotropies are deduced from the interface NMR signal.
\end{abstract}


\maketitle
\section{Introduction}
Recently, numerous experiments have revealed a profound effect of substrate on both transport and magnetic properties of epitaxial thin films La$_{0.67}$M$_{0.33}$MnO$_3$ (M=Sr, Ca, or Ba) which are ferromagnetic metals at low temperatures and paramagnetic insulators at high temperatures. In the manganite films, the substrate-induced strain affects magnetoresistivity,\cite{Jin,Lofland,Ziese} Curie temperature,\cite{Balestrino} and magnetic microstructure\cite{Lecoeur,Dho}. In particular, the magnetic anisotropy energy in the thin manganite films strongly depends on strain and, therefore, on the substrate material, film thickness, and deposition parameters.\cite{Steenbeck, SuzukiStrain,Berndt} For instance, whereas the La$_{0.67}$Ca$_{0.33}$MnO$_3$ films grown on SrTiO$_3$ substrate were found to have uniaxial magnetic anisotropy (hard axis/easy plane) with easy plane being the film plane, the possibility of producing strained manganite films deposited on LaAlO$_3$ substrate with the easy magnetization axis along film normal has been proposed.\cite{Odonnell} In addition, nuclear magnetic resonance (NMR) studies on a series of the epitaxial thin films discovered the existence of a complex phase separation phenomenon,\cite{Bibes} correlated to the insulating nature of the thinner films,\cite{Sun,Bibes1} that may reflect an intrinsic property of film/substrate interface. The material engineering potentials of this phenomenon, which  might appear at first sight a drawback for oxide materials, were recently demonstrated by an example of successful tailoring of this interface property.\cite{Yamada}

 It is therefore of crucial importance for future applications in tunneling magnetoresistive devices to characterize the origin of the strong film-substrate interactions. In this paper $^{55}$Mn NMR and magnetization data are used in order to correlate the microscopic and macroscopic properties of the manganite La$_{0.67}$Sr$_{0.33}$MnO$_3$/SrTiO$_3$ interfaces.
 The rest of the paper is divided into: Section II, on material preparation and experimental techniques; Section III,  on NMR and magnetization results, including discussion of a simple model; Section IV, with our conclusions. 

\section{Materials and Experiments}
Epitaxial La$_{0.67}$Sr$_{0.33}$MnO$_3$ (LSMO) films with thickness varying from 8 to 500 nm were grown on (100) SrTiO$_3$ (STO) substrates using the pulsed laser deposition technique. X-ray diffraction experiments show that the LSMO thin films exhibit a cube-on-cube type of epitaxial arrangement on the substrates. Deposition conditions, structural characterization and transport properties have been described elsewhere.\cite{Balestrino} It should be mentioned here that the  annealing duration for optimal oxygen content was found to vary with film thickness: two hours for thicker films and half an hour and less for thinner films.

For NMR experiments the home-built broadband fast-averaging spectrometer, HyReSpect,\cite{Allodi} was used. 
Two types of NMR  experiments are presented in this paper, both performed  at $T=1.6$ K. The zero-field spectra were obtained in the frequency range 300-450 MHz by means of a standard $90^\circ-\tau-90^\circ$ spin-echo pulse sequence, plotting point by point the amplitude at zero frequency shift of the Fast Fourier Transform of each echo as a function of transmission frequency. The plotted data are always corrected for the NMR sensitivity, dividing amplitudes by $\omega^2$. 

The field-swept NMR measurements were taken at fixed frequency with the same sequence in a static magnetic field $\bm{H}$ ranging from zero to 20 kOe, applied either parallel or perpendicular to the film surface. The radio frequency field $\bm{h}$ was parallel to the film plane, and always orthogonal to $\bm{H}$. In order to ensure that the sample was initially in a single domain state, we started from saturation and measured the NMR signal varying the magnetic field intensity $H$ along a full hysteresis loop.

NMR in ferromagnetic thin films exploits the so-called radio-frequency enhancement: an applied radio-frequency (rf) field $\bm{h}$ couples to the 
magnetization of a ferromagnetic material and tilts its electronic moments, thus producing a modulation of the hyperfine field at the nucleus at 
the same frequency.\cite{Turov,Portis} Since the hyperfine field is huge for transition metal nuclei, this gives rise to a large amplification 
(enhancement) of the rf field at the nucleus, 
$\bm{h} = \eta\bm{h}_0$.
The enhancement factor $\eta$ may be directly assessed by experiment from the optimal radio-frequency pulse excitation, which implies that:
\begin{equation}
2 \pi \gamma\eta h_0 \Delta t  = \pi/2 \label{etaexp}
\end{equation}
where $\gamma$ is the nuclear magnetogyric ratio ($\gamma=1.055$ MHz/kOe for $^{55}$Mn) and $\Delta t$ the pulse duration. The applied rf field $h$ may be calibrated by comparison with the optimal excitation condition for an equivalent non magnetic nucleus (i.e.\ {\em without} enhancement). 

Two mechanisms of enhancement may typically be distinguished in homogeneous ferromagnets. One is due to the rf-induced motion of the domain walls,\cite{Turov,Portis} it is generally dominant in zero field and obviously removed by the application of a static field approaching the saturation value. The other mechanism, which is usually less effective by one-two orders of magnitude, arises from the rotation of the magnetization in the bulk of domains, and it comes into play both in zero and in an applied field. The corresponding domain enhancement factor can be related to the  local field at the nucleus $\bm{H_n}$, and the effective magnetic anisotropy by a simple theoretical model.\cite{Turov}
Since $\bm{H_n}$ is essentially the hyperfine field (i.e.\ $H_n \gg H, h$)
\begin{equation}
\eta(H)=\frac{\partial{H}^{\bot}_{n}}{\partial{h}}=-H_{n}\text{sin}(\phi+\theta_0)\left( \frac{\partial{\theta_0}}{\partial{h}}\right) _{h=0}.\label{EnhFactor}
\end{equation}
Here ${H}^{\bot}_{n}$ is the component of the effective local field perpendicular to its direction for $h=0$, $\phi$ is angle 
between the easy magnetization axis and $\bm{h}$, and $\theta_0$ is the angle between the easy axis and the electronic  equilibrium magnetization $\bm{M}$ in a magnetically anisotropic sample, which may be found by minimizing the total magnetic energy.  

Axial magnetization was measured by using a commercial superconducting quantum interference device (SQUID), 5 T Quantum MPMS, at $T=2$ K and in the 0 - 30 kOe field range, both with the field parallel and perpendicular to the film surface. We avoided spurious contributions to the measured signal, other than from the film and the substrate, by using diamagnetic sample holders with uniform mass and magnetic moment distribution along the whole SQUID scanning length. The magnetization data of the manganite films were corrected for the observed diamagnetic contribution of the substrate. The maximum fields misalignment in both NMR and SQUID measurements was estimated to be of the order of five degrees.

\section{Results and Discussion}
\subsection{Zero-field $^{55}$Mn NMR}
In the mixed valence metallic manganites different manganese states yield distinct contributions to the NMR spectra. The localized Mn$^{4+}$ state gives rise to a peak between 310 and 330 MHz.\cite{Matsu, Allodi1, Kapusta, Savosta2} The localized  Mn$^{3+}$ resonance strongly depends on the local spin and orbital directions. Its spectral position was shown to vary between 250 MHz and 450 MHz in a Mn ferrimagnetic spinel, in which the spin orientation could be controlled experimentally.\cite{KuboSpinel} The peculiar orbital and spin structure of pseudocubic manganites  confines the Mn$^{3+}$ in the range  350-430 MHz, easily distinguished from the Mn$^{4+}$ contribution.\cite{Matsu, Allodi1, Kapusta, Savosta2} Finally, the signal corresponding to the Mn$^{DE}$ state from the mixed valence metallic region is associated with a fast hopping of electrons among Mn sites and it shows up  as a relatively narrow peak at an intermediate frequency in the 370-400 MHz range.
\begin{figure}
\includegraphics[scale=0.4]{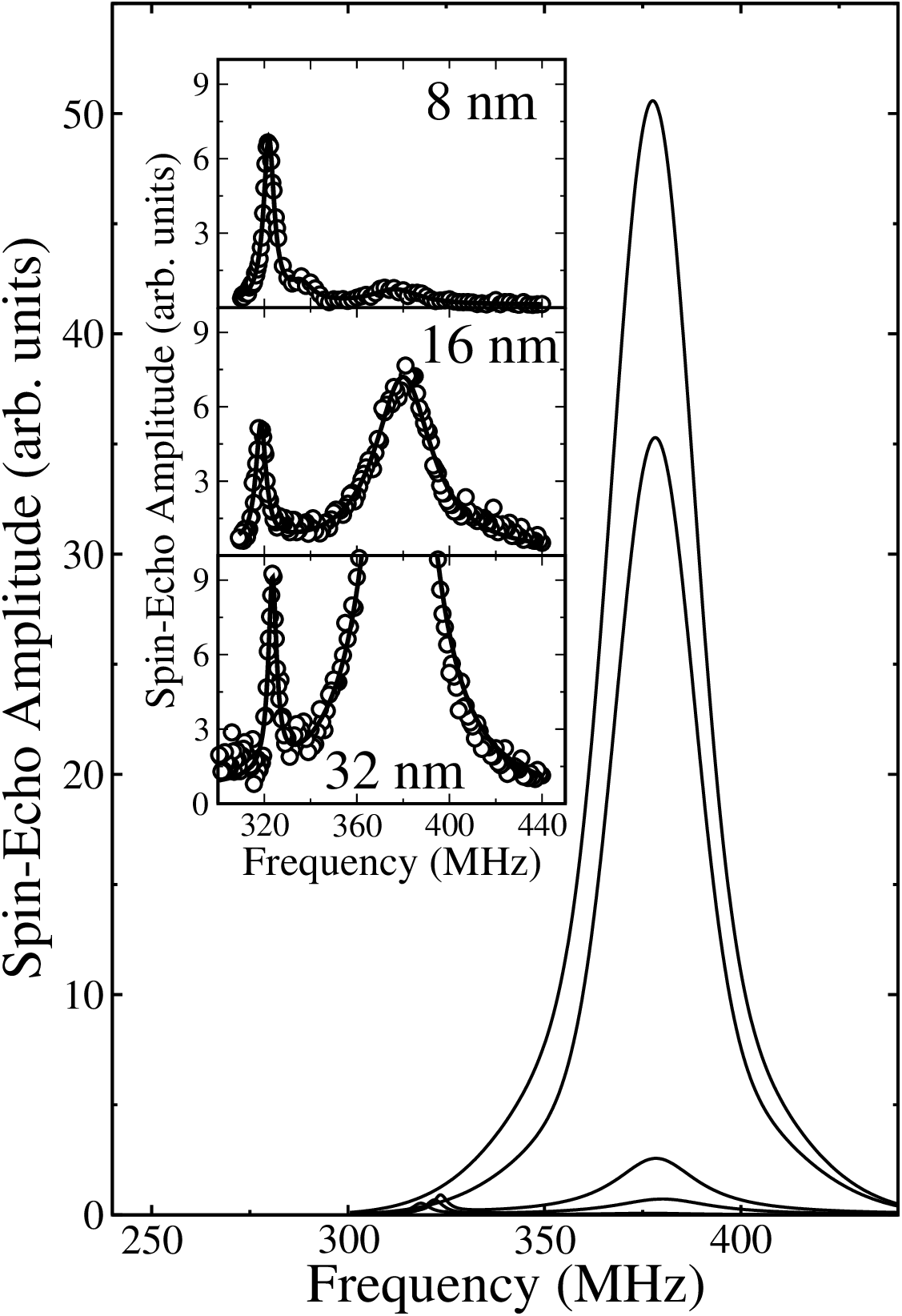}
\caption{\label{fig:PLSMOspec} Zero-field $^{55}$Mn spin-echo amplitude measured at 1.6 K in LSMO/STO films with different thickness.}
\end{figure}

Zero-field $^{55}$Mn NMR spectra obtained at 1.6 K in LSMO/STO films with various thickness are shown in Fig.~\ref{fig:PLSMOspec}. Two distinct lines at $f^{4+}\approx322$ MHz and $f^{DE}\approx380$ MHz, corresponding to two different phases with the localized charges (Mn$^{4+}$ state) and with itinerant carriers (Mn$^{DE}$ state), respectively, are observed. Their origin was already assigned by previous NMR work\cite{Bibes}  on La$_{0.67}$Ca$_{0.33}$MnO$_3$ to the inhomogeneous separation of a so-called {\em dead-layer}, roughly coinciding with the critical thickness below which the same films appear to be insulating by transport measurements.  The dead-layer is presumably located at the interface between the substrate and the manganite film. The absence of a corresponding localized Mn$^{3+}$ signal may be attributed to its much faster relaxation rate and possibly to its wider spectral breadth.
It should be noted that detailed studies of ultrafine LSMO particles have also revealed the existence of a small contribution from Mn$^{4+}$ state, attributed to the surface of the nanoparticles.\cite{Savosta} Also in that case the corresponding Mn$^{3+}$ signal was missing.

\begin{figure}
\includegraphics[scale=0.4]{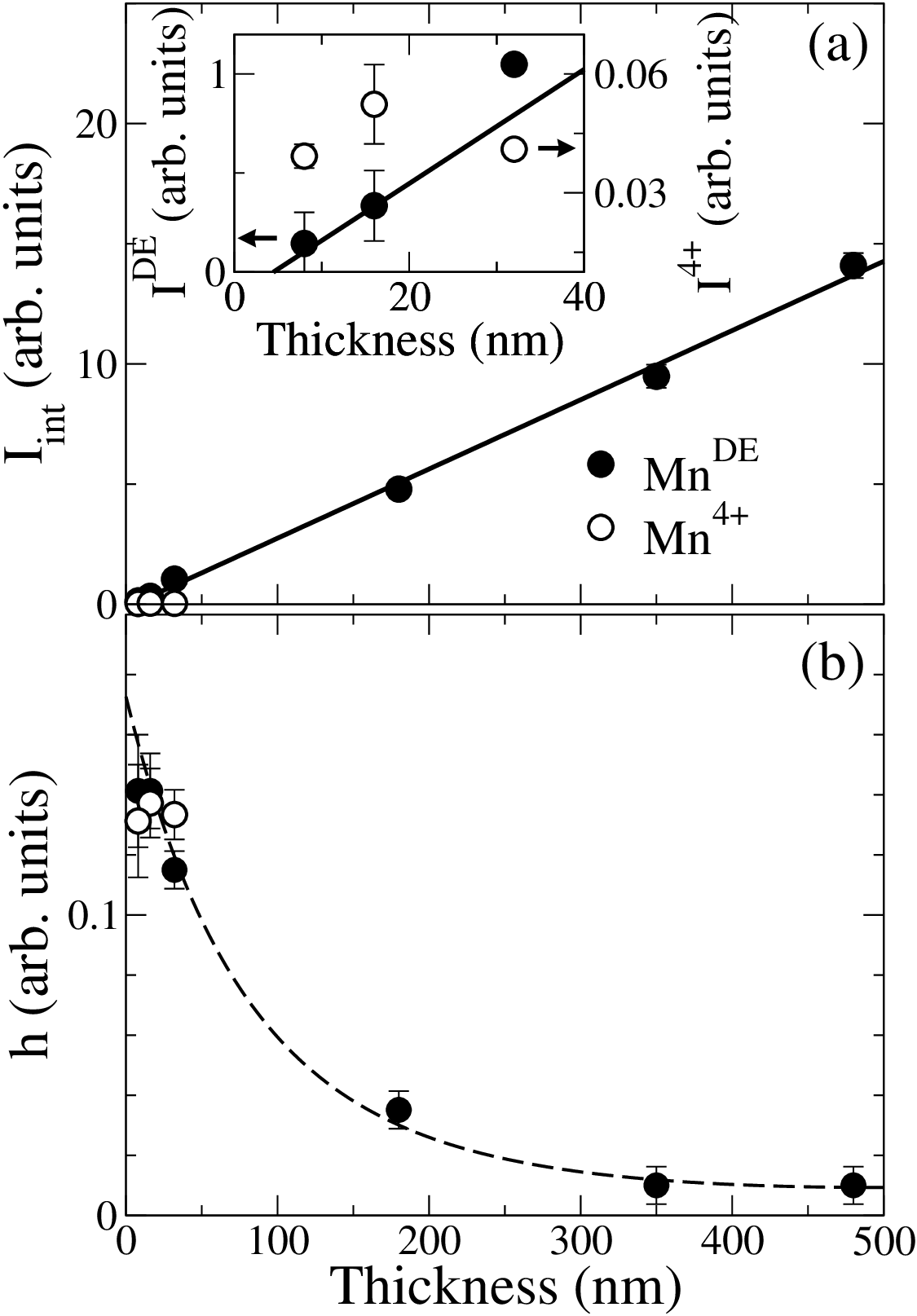}
\caption{\label{fig:PLSMOarea}Thickness dependence of: (a) the integrated intensity I$_{int}$ of the $f^{DE}$ and of the  $f^{4+}$ lines; (b) the applied rf field $h$, proportional to $\eta$ (see text). Inset: Zoom on thiner films.}
\end{figure}

As can be seen the peak intensity at $f^{DE}$ dramatically depends on the film thickness $\textit{t}$, whereas the intensity of the peak at $f^{4+}$ is practically unvaried, and, in films with thickness $t\gtrsim180$ nm, a separate line at $f^{4+}$ is difficult to identify due to overlapping with the strong signal from Mn$^{DE}$. The values of the integrated spectrum intensity obtained for the $8\leq t \leq480$-nm films are plotted as a function of $\textit{t}$ in Fig.~\ref{fig:PLSMOarea}(a); within the experimental errors the integrated intensity follows a linear dependence on $t$, intersecting the horizontal axis at $t_{d}\approx4.6\pm 0.5$ nm, as can be seen in the inset. The obtained $t_d$ value is therefore interpreted as the dead-layer thickness and it is comparable  with the value $5.3$ nm  determined  on La$_{0.67}$Ca$_{0.33}$MnO$_3$.\cite{Bibes}
 Values of the critical thickness were measured by resistivity in LSMO films on different substrates, yielding comparable results: $\sim 3$ nm on (001) LaAlO$_3$  and $\sim5$ nm on (110) NdGaO$_3$.\cite{Sun} Similar measurements were performed on our films; they confirm the presence of a critical thickness of this order of magnitude.\cite{Balestrino}

Figure~\ref{fig:PLSMOarea}(b) shows the thickness dependence of the applied rf field $h$ required to optimally excite the  $f^{DE}$ NMR resonance. The reduction of $h$ with increasing film thickness indicates an increase of the experimental enhancement factor $\eta=h/h_0$, as given by Eq. \ref{etaexp}.  

In contrast to the NMR study on La$_{0.67}$Ca$_{0.33}$MnO$_3$ thin films,\cite{Bibes} a monotonic shift of the $f^{DE}$ peak towards lower frequency with reduction of the film thickness was not observed for LSMO.

\subsection{Field sweep spectra}

We shall start by considering the field dependence of the $f^{DE}$ line of  the 180-nm thick LSMO/STO film, arising from the metallic region. The experiments were run in two distinct geometries, with the external dc field either in the film plane ($H_\parallel$) or perpendicular to it ($H_\perp$), and with the rf field $h$ always in the film plane, at right angles with the dc field.  

The field dependent resonance frequency is defined as that corresponding to the maximum amplitude, as determined from the best fit of the Mn$^{DE}$ spectrum to a Lorentzian lineshape. 

\begin{figure}[htb]
\includegraphics[scale=0.3]{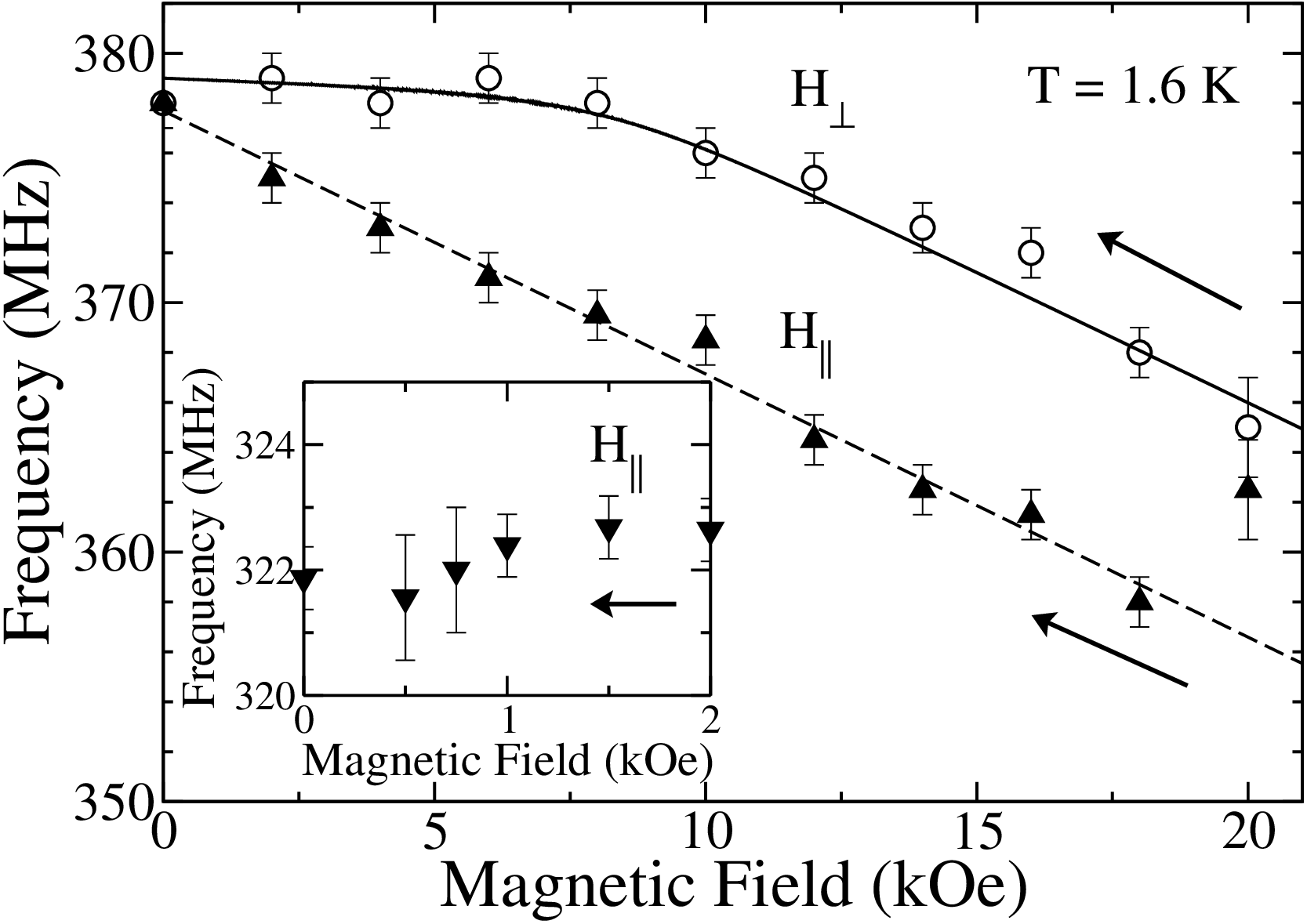}
\caption{\label{fig:FreqShift} $^{55}$Mn nuclear resonance frequency $f^{DE}$ in the 180-nm thick film as functions of the dc field in the film plane ($\blacktriangle$) and perpendicular to the film plane ($\circ$).  The solid and dashed lines are calculated curves in the single domain model (see text). The inset shows the shift of the $f^{4+}$ resonance line in the 8-nm thick film measured with the field applied in the film plane at 1.6 K.}
\end{figure}

Figure~\ref{fig:FreqShift} shows the shift of the $f^{DE}$ line measured at various dc field up to 20 kOe in both geometries, at T$\approx$1.6 K. In the $H_\parallel$ geometry (filled symbols) the resonance spectrum shifts to lower frequencies at a rate close to the Mn magnetogyric ratio $\gamma$. The negative slope indicates that the hyperfine field is negative,\cite{Freeman} that is, it lies antiparallel to the electronic magnetization. 
When the field is applied perpendicular to the film plane, the demagnetizing field also contributes to the shift in the resonance frequency and in the NMR intensity. As the magnetic field is increased, the demagnetizing field increases up to $4\pi M_s$, and the resonance frequency remains almost independent on the field up to $\sim 8$ kOe (Fig.~\ref{fig:FreqShift}, open circles).

Let us now turn to the $f^{DE}$ echo intensities versus field at fixed frequency. 
We recall that in homogeneous ferromagnets domain bulk and domain walls give rise to NMR signals distinguished by very different enhancement factors. In low fields the main contribution to the spin-echo intensity is given by the nuclei within the domain walls and its amplitude is expected to decrease with increasing external magnetic field.  
In intrinsically inhomogeneous materials like the manganites it is not possible to experimentally separate domain bulk and domain wall signals by distinguishing their enhancement factors.  However, in order to observe a dominant domain  wall contribution the duration and the power of the rf pulses may be  fixed at values optimized in zero dc field (ZFO), whereas if both the duration and the power of the rf pulses are adjusted for  maximum response at each field (FO, field optimized), a signal is observed also above the effective anisotropy field, where the sample is saturated and the NMR  must originate from the bulk of the single domain.

\begin{figure}
\includegraphics[scale=0.3]{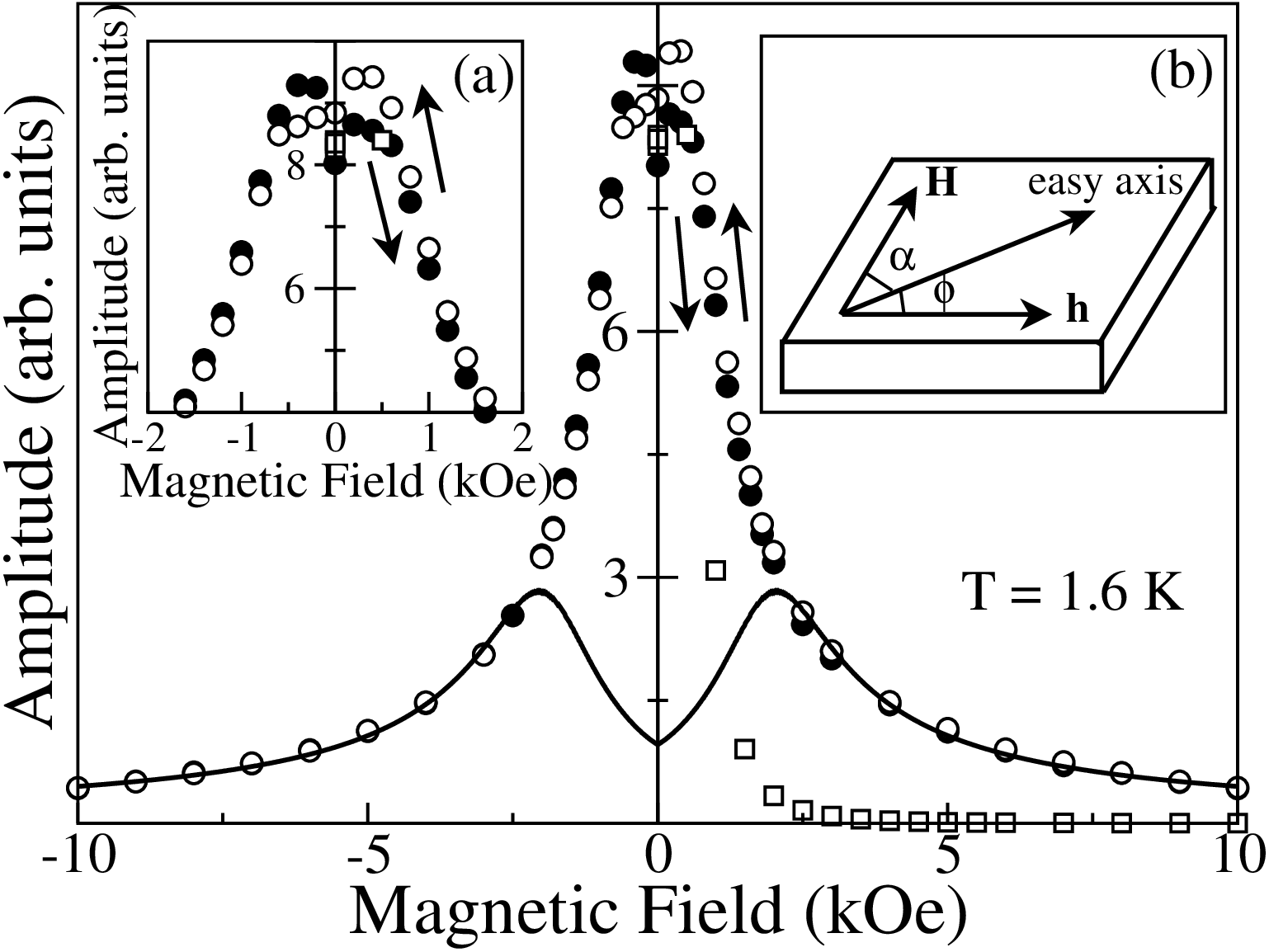}
\caption{\label{fig:AmplinPlane}Field dependence of the spin-echo amplitude at 1.6 K, $f=380$ MHz, for the 180 nm thick film, both with the FO (o decreasing fields, $\bullet$ increasing fields) and the ZFO ($\square$) protocol. The rf field and the dc field directed along the hard magnetization axes. The solid line represents the calculated curve in the single domain model (see text). Insets: (a) hysteresis of the NMR amplitude in the vicinity of the zero field; (b) Orientations of $\bm{H}$ and $\bm{h}$ with respect to the easy axis,}
\end{figure}

Figure~\ref{fig:AmplinPlane} displays the field dependences of the spin-echo intensity measured at 1.6 K, $f=380$ MHz, according to the two above mentioned protocols, for the dc magnetic field applied in the film plane along the hard axis [001].  Both variations of the experiment produce a dramatic reduction of amplitude with field. Since the linewidth (Fig.~\ref{fig:PLSMOspec}) is quite larger than the shifts (Fig.~\ref{fig:FreqShift}) this amplitude drop is not due to a shift of spectral weight outside the experimental passband. 
The open squares refer to the ZFO protocol and the signal thus obtained disappears with the application of modest magnetic fields, as domain walls are removed.
Assuming that the domain wall enhancement does not vary dramatically with field, the field dependence of the ZFO signal amplitude reflects approximately the reduction of the domain wall volume. 

The FO signal amplitudes were recorded starting from $H_{max}=25$ kOe, in order to ensure a saturated sample, sweeping down the field (open circles) through zero to  $-H_{max}$, and then reversing the field sweep direction (closed circles).  
Notice that  in the vicinity of zero field a hysteretic behavior of the amplitude is observed (see the inset (a) of Fig.~\ref{fig:AmplinPlane} ). The measured spin-echo amplitude peaks at $\pm 200(50)$ Oe, a value which corresponds to the coercive filed $H_c$, in agreement with our magnetization data (not shown) and with the literature\cite{Steren}.

\begin{figure}
\includegraphics[scale=0.3]{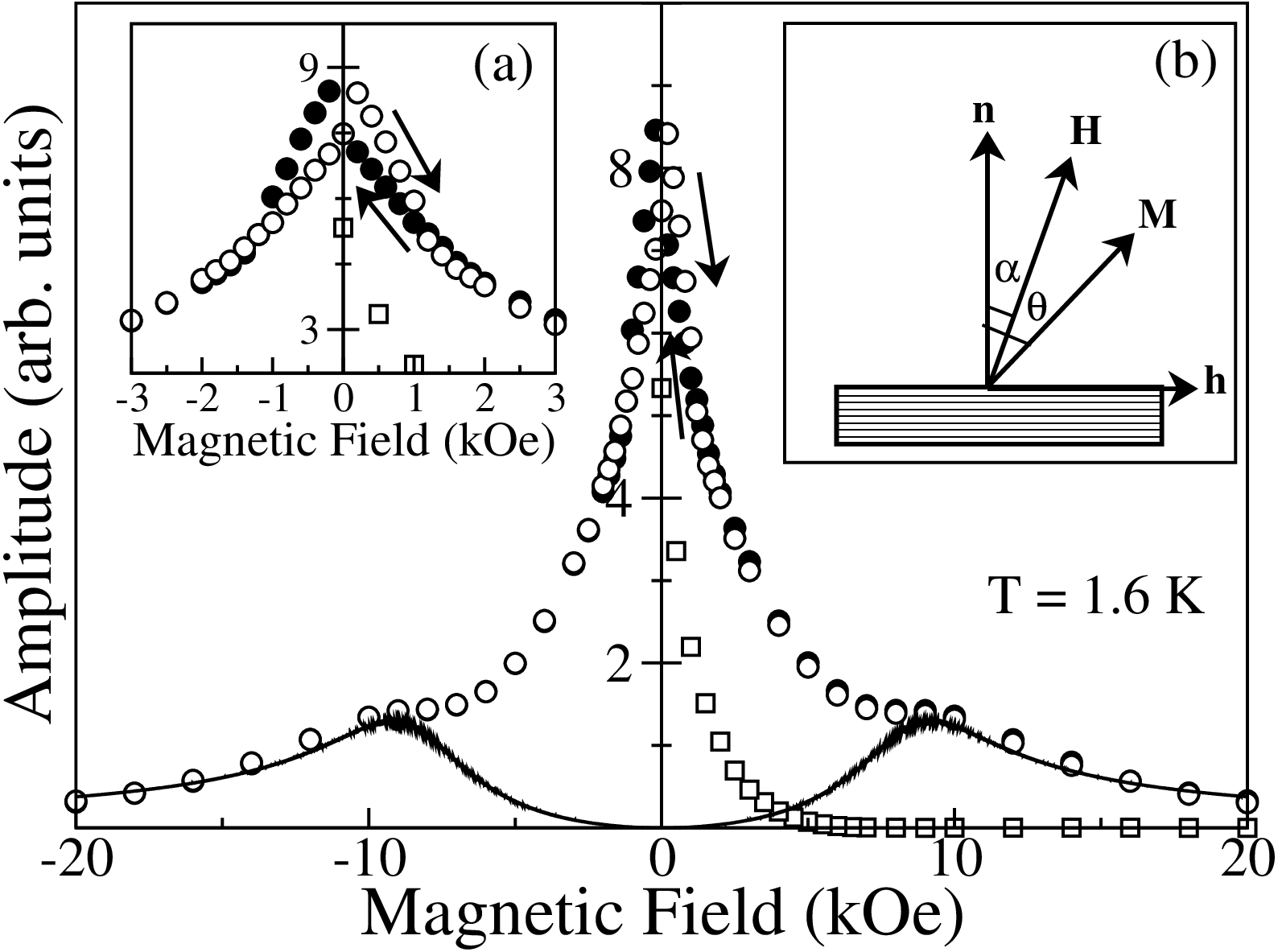}
\caption{\label{fig:AoutPlane}Field dependence of the spin-echo amplitude at 1.6 K, $f=380$ MHz, for the 180 nm thick film, both with the FO (o decreasing fields, $\bullet$ increasing fields) and the ZFO ($\square$) protocol. The dc field is nearly perpendicular to the film plane. The solid line represents the calculated curve in the single domain model (see text). Insets: (a) The hysteresis of the NMR amplitude in the vicinity of the zero field; (b) Orientations, with respect to the film normal, of $\bm{H}$ and $\bm{M}_s$.} 
\end{figure}

NMR $f^{DE}$ amplitudes were recorded according to both ZFO and FO protocols also for the $H_\perp$ geometry at $f=380$ MHz (Fig.~\ref{fig:AoutPlane}). A behavior similar to that of Fig.~\ref{fig:AmplinPlane} is obtained for the ZFO curves, whereas the FO signal displays a marked feature around $H\approx 10$ kOe, which roughly corresponds to the demagnetizing field.

\begin{figure}[!b]
\includegraphics[scale=0.3]{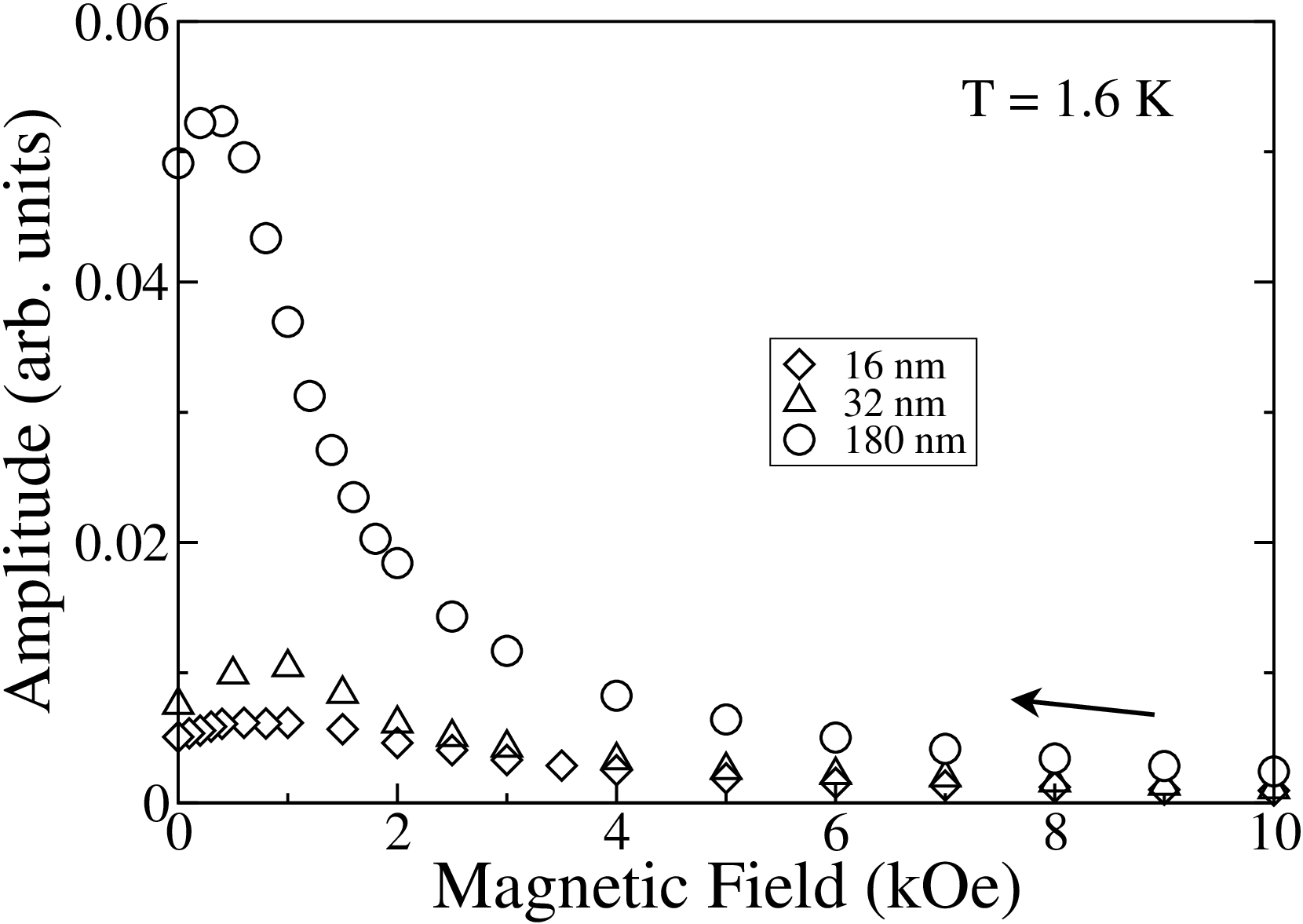}
\caption{\label{fig:AmplinPlaneAll} Field dependence of the spin-echo amplitude at 1.6 K of the manganite films with different thickness.}
\end{figure}

It should be noted that the all our films revealed very similar behavior of the spin-echo amplitude versus external magnetic field:  Fig.~\ref{fig:AmplinPlaneAll} shows the FO intensities for two thinner films (32 and 16 nm) together with the 180 nm results of Fig.~\ref{fig:AmplinPlane}. 

\begin{figure}
\includegraphics*[scale=0.3]{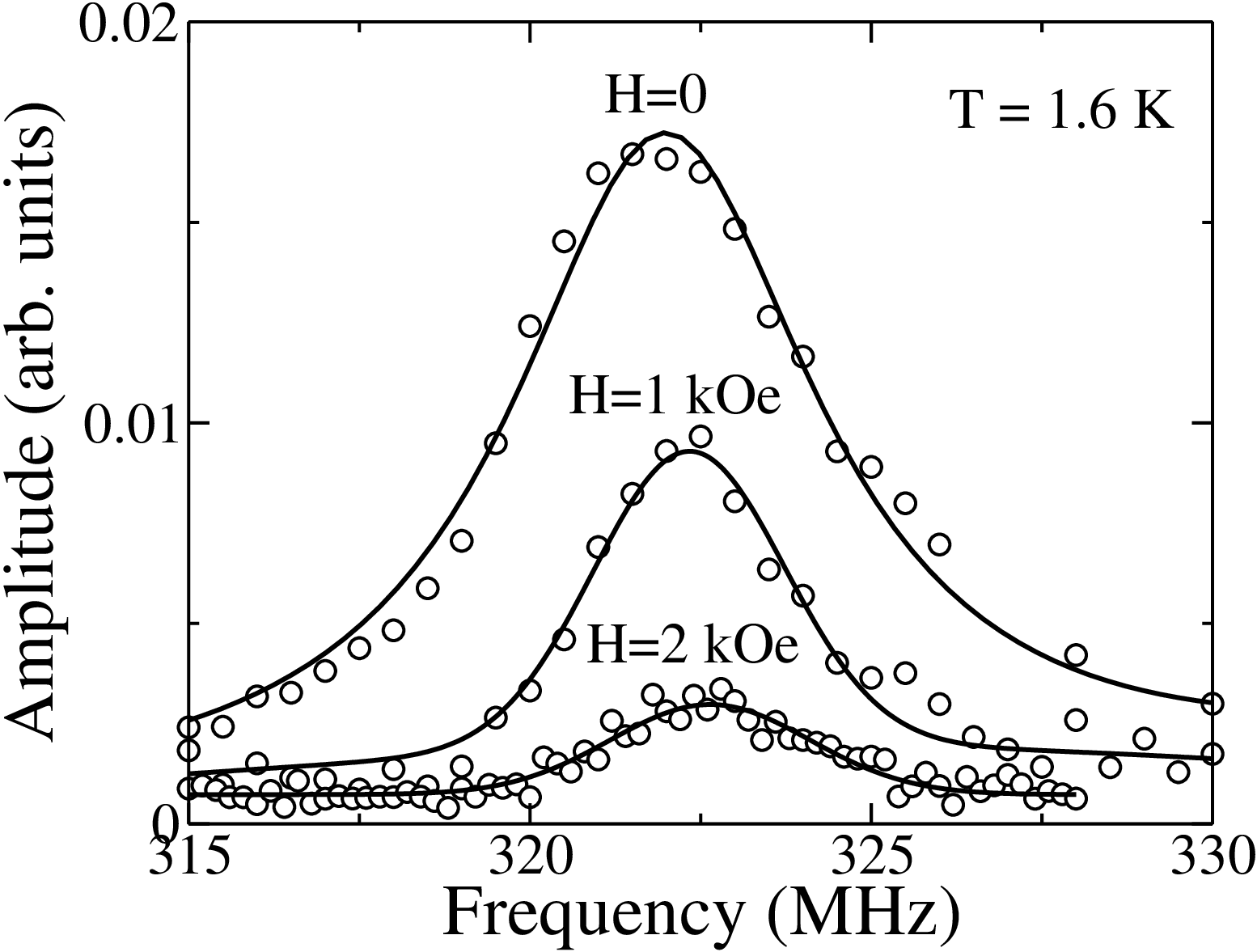}
\caption{\label{fig:PLSMOspecMn4} $^{55}$Mn spin-echo spectra at 1.6 K, corresponding to the insulating region of the LSMO/STO film with thickness of 8 nm, for different values of the dc field applied in the film plane.}
\end{figure}

We turn now to the field dependence of the $f^{4+}$ line, corresponding to the insulating region of the film. Figure~\ref{fig:PLSMOspecMn4} represents an evolution of the spin-echo spectra, measured at 1.6 K on the LSMO/STO film with thickness of 8 nm. The three spectra refer to different values of the dc field applied in the film plane. The FO and ZFO protocols coincide for this signal, i.e. the optimal irradiation conditions do not depend on the applied dc field. Furthermore the enhancement factor, as determined from Eq.~\ref{etaexp}, is comparable with that of the thinner film $f^{DE}$ line (see Fig.~\ref{fig:PLSMOarea}(b)).

The NMR signal at $f^{4+}\approx322$ MHz disappears already in quite weak fields $\sim3$ kOe. Since $\eta(H)$ for a single domain drops hyperbolically with field (see Eq.~\ref{Amplitude} below), the lack of a FO signal at high magnetic fields could just be due to insufficient NMR sensitivity for the very small volume of the dead-layer. Viceversa, the Mn$^{4+}$ signal is observed in zero dc field thanks to a large value of $\eta(H=0)$, indicative of the ferromagnetic nature of the insulating region. Notice that the spectra of Fig.~\ref{fig:PLSMOspecMn4} do not shift appreciably with the applied field, as it is further confirmed by the inset of Fig.~\ref{fig:FreqShift}.

\subsection{Single domain model}

\begin{figure}
\includegraphics[scale=0.3]{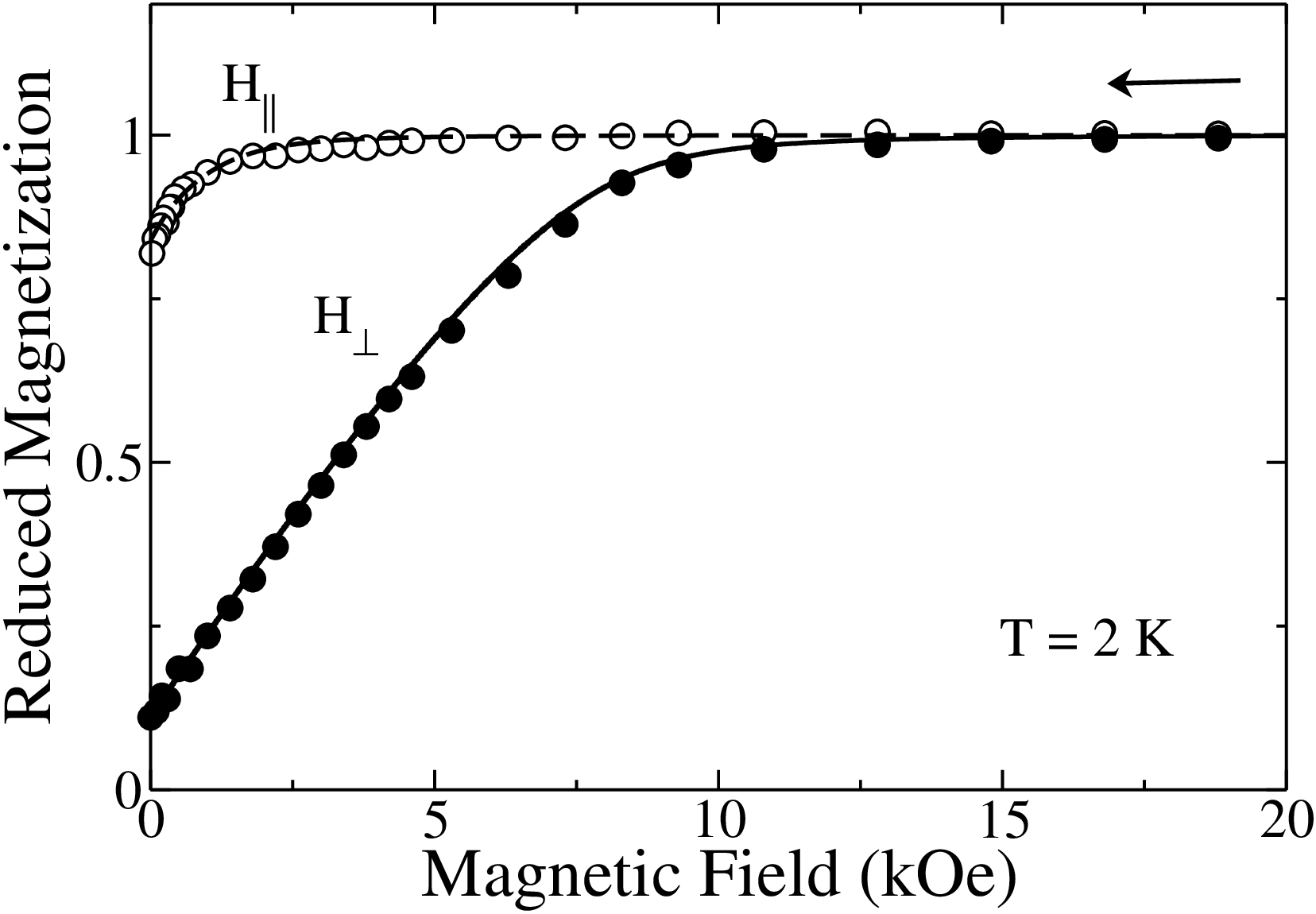}
\caption{\label{fig:MagnOutPlane}Field dependence of the reduced in-plane (o) and out-of-plane ($\bullet$) magnetization at 2 K of the 180 nm thick film. The solid and dashed lines are calculated curves (see text).}
\end{figure}

For large enough applied fields we can describe our films in a simple single domain model. We can further assume a biaxial magnetic anisotropy in the (100) film plane,\cite{Lecoeur, SuzukiStrain} with easy axis along [110], as it was directly checked by SQUID on our film.  Hence the total energy for an ideal cubic crystal is given by:
\begin{equation}
E=K_{1}\sin^{2}\theta\cos^{2}\theta-HM_{s}\cos(\alpha-\theta)-hM_{s}\cos(\phi+\theta)\label{Energy}
\end{equation}
where  $\theta$, $\phi$, and $\alpha$ are angles between the easy magnetization axis and the vectors $\bm{M}_s$, $\bm{h}$, and $\bm{H}$, respectively, as shown in Fig.~\ref{fig:AmplinPlane}~(b).
Then for a resonant excitation of the nuclei in the domain bulk, Eq.~(\ref{EnhFactor}) predicts the following field dependence for the echo amplitude:\cite{Turov}
\begin{equation}
A(H)=\text{const}\frac{\sin^{2}(\phi+\theta_{0})}{(2K_{1}/M_{s})\cos(4\theta_0)+H\cos(\alpha-\theta_{0})}\label{Amplitude}
\end{equation}
where we recall that $\theta_0$ is the  angular coordinate of the equilibrium magnetization. In order to compare this expression with the experimental data of Fig.~\ref{fig:AmplinPlane} we need to determine the value of the effective anisotropy field $H_{K}=2K_{1}/M_{s}$. To this end we employed SQUID on the same film and measured the reduced magnetization $M/M_s$ shown in Fig.~\ref{fig:MagnOutPlane}, where the saturation magnetization value $M_s=620$ emu/cm$^3$ was determined at 2 K.
The reduced magnetization may be fitted to the expression $M/M_s=\cos( \theta_0 - \alpha)$, where the equilibrium angle $\theta_0$ is found by minimizing the energy of Eq.~\ref{Energy} with $h=0$. The dashed line in Fig.~\ref{fig:MagnOutPlane} represents the best fit, which corresponds to a value of $H_{K}=1.7$ kOe,  
that gives an effective in-plane anisotropy constant $K_{1}\approx 52.9\cdot 10^4$ erg/cm$^3$, together with $\alpha=40^\circ$, 
consistent within the accuracy of sample alignment
with the easy magnetization axis along [110] . 

The solid curve in Fig.~\ref{fig:AmplinPlane} corresponds to the prediction of Eq.~\ref{Energy} and Eq.~\ref{Amplitude}, for the quoted value of the effective anisotropy field $H_K$ with $\alpha=40^\circ$, $\phi=50^\circ$. The curve agrees well with the data for fields in excess of $H_K$, where a single domain structure is expected, and the maximum of $A(H)$ corresponds to the  anisotropy field $H_K$. 
At $H < H_K$, the data deviate from the theoretical curve due to the dominant 
contribution of the domain-wall signal, which could not be unraveled from the 
signal of domains.   
Also the field dependence of the resonance frequency of Fig.~\ref{fig:FreqShift} may be calculated without further adjustable parameters from the same minimization, by means of the projection of the magnetic field onto the local field, $\delta H = - H \cos(\theta_0-\alpha))$ (dashed curve in the same figure).

The overall behavior of this single domain model in the magnetic field may be qualitatively understood as follows:  
Eq.~\ref{EnhFactor} states that the enhanced NMR signal is proportional to $\frac{\partial{H^\perp_n}}{\partial{h}}$, where $H^\perp_{n} = (M^\perp/M_s)H_{n}=\cos(\theta_0-\alpha) H_{n}$. 
In zero field, the magnetization is aligned along the easy axis [110] and so $\cos(\theta_0-\alpha)=1/\sqrt 2$. When the external dc field $H$ is increased in the geometry of Fig.~\ref{fig:AmplinPlane} (b), the angle $(\theta_0-\alpha)$ tends to zero as the magnetization $\bm{M}$ rotates towards the dc field direction. As consequence, the amplitude of the NMR signal grows until magnetic saturation is reached at $H=H_K$. The  NMR amplitude reduction at larger fields is due to the fact that the radiofrequency driven deviation of  $\bm{M}$ (hence of the effective local nuclear field $\bm{H}_n$) from the equilibrium position is becoming smaller and smaller as $H$ increases.

 We now turn to the  field dependence of the resonance frequency in the $H_\perp$ geometry where we neglect the small in-plane magnetic anisotropy, so that the dc magnetic field $\bm{H}$, the magnetization $\bm{M}$, and the film normal $\bm{n}$ are coplanar. Thus the total energy is given by expression:
\begin{equation}
E=-HM_{s}\cos(\theta-\alpha)+K_u \sin^{2}\theta+2\pi M^{2}_{s}\cos^{2}\theta \label{MagEnergyTrans}
\end{equation}
where $\alpha$ and $\theta$ are the angles of $\bm{H}$ and $\bm{M}_s$ from the film normal, respectively; $K_{u}$ is an effective out-of-plane uniaxial anisotropy constant. The shift in the resonance field $\delta H$ is given by projection of the magnetic and demagnetizing field onto the local field $H_n$ at the Mn nuclei:
\begin{equation}
\delta{H}=4\pi M_{s}\cos^2 \theta_0-H\cos (\theta_0-\alpha) \label{Shift}
\end{equation}
The best fit of the reduced magnetization of Fig.~\ref{fig:MagnOutPlane} to the expression $M/M_s=\cos(\theta_0-\alpha)$, where $\theta_0$ is obtained minimizing Eq.~\ref{MagEnergyTrans}, yields an effective anisotropy field value $H_A=4\pi M_s+2K_u/M_s= 7.86$ kOe.  The term $2K_u/M_s$ represents a possible perpendicular uniaxial anisotropy, and, since we have measured the demagnetizing field $4\pi M_s=7.79$ kOe by SQUID, the fit provides an estimated value $2K_u/M_s=70$ Oe, corresponding to $K_u\le2.2\cdot10^4$ erg/cm$^3$, i.e. we may deduce the absence of a significant anisotropy with a symmetry axis normal to the plane of the film.

The shift of the resonance frequency was calculated from Eq.~\ref{Shift} using this $H_A$ value. The solid curve in Fig.~\ref{fig:FreqShift} represents this calculation where the only adjustable parameter is the estimated misorientation $\alpha=5^\circ$. The agreement with the data is quite good. The same model predicts the field dependence of the  FO NMR intensity  displayed Fig.~\ref{fig:AoutPlane} (circles) by means of Eq.~\ref{EnhFactor}. Here too a good agreement is found for fields in excess of $H_A$.

\section{Conclusion}
$^{55}$Mn NMR detects two distinct signals from localized holes in the LSMO/STO interfaces and from the itinerant carriers of the upper layers of the films. The investigation of NMR spectra vs. film thickness confirmed the presence of a dead-layer (non metallic) at the interface with the substrate. 

Although our main NMR result on LSMO/STO confirms qualitatively the findings of Bibes {\em et al.}\cite{Bibes} on LCMO/STO, we notice two subtler differences. We do not detect evidence of an additional  {\em non magnetic} insulating region at the interface, which was inferred from the thickness dependence of the $f^{4+}$ intensity in LCMO/STO. Our LSMO/STO $f^{4+}$ intensity is thickness-independent.
The Mn$^{DE}$ frequency value shifts linearly \cite{Allodi1} with charge carrier density and the variation with LCMO film thickness was attributed \cite{Bibes} to a corresponding dependence of the average density of carriers. We do not detect a similar change of the Mn$^{DE}$ resonance frequency in LSMO/STO.  

These differences may be related to specific properties of LCMO and LSMO, or else they may be related to sample preparation conditions. In both cases they indicate that NMR, in conjunction with magnetization measurements, is a very sensitive tool of the interface quality.

By examining the high field region of the field dependence of the NMR of LSMO/STO interfaces, the NMR response can be fitted to a simple model that allows independent determination of the anisotropy field in the films. Thus our results reveal agreement between the NMR and magnetization measurements.
Specifically a comparison of the calculated and experimental  field dependence of the  $^{55}$Mn NMR and SQUID results shows that magnetization rotation processes play a dominant role when the applied field exceeds the effective anisotropy field. 

The marked difference in frequency shift between $f^{4+}$ and $f^{DE}$, 
displayed in Fig,~\ref{fig:FreqShift} indicates that the layers of the manganite films located close to the substrate are magnetically highly anisotropic, whereas the upper film layers are only slightly anisotropic. This observation is consistent with the dependence of the rf power on film thickness, namely that in order to excite nuclei in the thiner films a larger radio frequency field is required, as it is expected in the presence of a larger magnetic anisotropy.

In conclusion we have demonstrated that $^{55}$Mn NMR can be further exploited for probing the magnetic properties of films in an interface-selective way, which is a key issue of the design of spintronic junction devices.

\section*{ACKNOWLEDGMENT}
This work was supported by the PRIN project (Grant No. 2002023998) and by the FIRB project (Grant No. RBNE017XSW).
\bibliography{Films}
\end{document}